\documentclass[%
 reprint,
nofootinbib,
 amsmath,amssymb,
 aps,
]{revtex4-2}

\usepackage{graphicx}
\usepackage{dcolumn}
\usepackage{bm}

\usepackage{braket}

\begin{document}

\preprint{APS/123-QED}

\title{Improved State Readout in NV Centers using \\
Regression Models and Rabi Driving}

\author{Fritz Haltenberger}
 \email{Corresponding author: haltenberger@em.uni-frankfurt.de}
\author{Manpreet Singh Jattana}%
 \email{jattana@em.uni-frankfurt.de}
\affiliation{%
 Modular Supercomputing and Quantum Computing,\\
 Goethe University Frankfurt, Kettenhofweg 139\\
 60325 Frankfurt am Main, Germany}%


\date{\today}


\begin{abstract}
Readout of state populations in nitrogen-vacancy centers from fluorescence measurements at room-temperature is routinely achieved via contrast-based calibration. The fidelities achieved by this conventional approach are limited by reducing the dynamical fluorescence behaviour of the NV center to a scalar value, and calculating the population of each possible state independently. To address these limitations, we use regression models trained on experimental data to map the fluorescence signals onto ideal simulated populations. Additionally, we enhance the informational content of the fluorescence signals by performing measurements during induced Rabi oscillations. Our results demonstrate that including these dynamical signals significantly reduces state readout errors across multiple tested models. Notably, linear ridge regression performs nearly on par with a non-linear kernel-based model, showing that simple models already capture the relevant mapping between the enhanced fluorescence signals and the underlying state populations. This data-driven approach provides a robust alternative that achieves higher fidelities than conventional calibration in our setting, paving the way for high-fidelity state readout in solid-state quantum registers. 
\end{abstract}

\maketitle


\section{Introduction}
Nitrogen vacancy ($NV$) centers in diamond are an interesting platform for quantum sensing \cite{MAZE_SENSING2008} and quantum computing \cite{DOHERTY20131, CHILDRESS_REVIEW2013}. Long coherence times of the electron spin \cite{GAEBEL_2006, JELEZKO_ELECTRON_2004} and nuclear spins \cite{PFENDER_SPECTRO_NUCLEAR_2019} of the negatively charged NV center ($NV^{-}$) allow coherent quantum operations even at room temperature \cite{NEUMANN_ROOMT_2008, VANOORT_ODMR_2000, JATTANA_2025}. This removes the need for cryogenic infrastructure, which can be expensive, difficult to maintain, and therefore limits access to well-funded or highly specialized laboratories. Consequently, NV-center-based platforms significantly lower the entry barrier for experimental quantum technologies and enable a broader range of research groups to participate in their development. 

We compare multiple methods' capabilities to infer the quantum state populations of a two-qubit system within an NV center in a synthetic diamond. The negatively charged NV center hosts two unpaired electrons, which form a spin-triplet ground state ($S=1$), with three magnetic sublevels ($m_s\in\{-1,0,1\}$) \cite{DAVIES_1994}. Following the standard convention \cite{HE_2024}, the computational $\ket{0}$ state of the first qubit corresponds to the $\ket{m_s=0}$ state of this electronic spin, and the computational $\ket{1}$ corresponds to the $\ket{m_s=-1}$ spin state. The second computational qubit is encoded in the nuclear spin of the $^{14}N$ atom of the NV center, where the $\ket{\uparrow}$ state will be used as the computational $\ket{0}$, and the $\ket{\downarrow}$ state as $\ket{1}$. Together, these two spins form the computational basis \{$\ket{00}$, $\ket{01}$, $\ket{10}$, $\ket{11}$\}. 


Optical readout of qubits in an NV-center-based quantum computer is achieved by measuring the spin-dependent fluorescence of the NV center \cite{GRUBER_1997, GULKA_2021}. One measures how many photons are emitted by the NV center while it is excited by a laser, which induces repeated optical excitations and radiative decay of its electrons \cite{CHILDRESS_2006}. The statistics of this fluorescence and the number of emitted photons is determined by the optical transitions and energy differences between the electronic states of the system. Importantly, different spin configurations of the system lead to different optical transitions and therefore distinguishable photon count statistics \cite{ROBLEDO_2011, MANSON_2006}. At cryogenic temperatures, even though only a fraction of the emitted photons is detected due to limited collection efficiency of the optical setup, the photon count statistics of the two electronic spin configurations that are used for quantum computation barely overlap. Therefore, for a given number of emitted photons within a single measurement process, one can infer with high confidence the electronic spin state \cite{ROBLEDO_2011}. 

At room temperature, however, phonon interactions lead to broader spectra of optical transitions, and increase the overlap between the photon count distributions of the two relevant electronic spin states \cite{PANADERO_2024, SCHMUNK_2012}. Therefore, the number of measured photons in a single measurement process contains significantly less information about the spin state of the electron, making \textit{single-shot readout} difficult at room temperature. Instead, the measurement process has to be repeated many times in order to obtain sufficient photon statistics \cite{STEINER_2010, JIANG_2009}. The aim of our work is to achieve higher fidelities for room-temperature readout of NV-center qubit registers by developing improved methods for inferring quantum state populations from repeated optical measurements.


We collect a dataset of structured photoluminescence measurement signals from a variety of quantum circuits, for which we know the expected noiseless populations. On this dataset, we benchmark different models' capacity to infer the expected populations from the measurement signals, and find that data-driven approaches outperform state-of-the-art contrast methods. 
In addition, we show how adding dynamical information to the measurement signals, by using controlled Rabi oscillations \cite{JELEZKO_ELECTRON_2004} of the final state, improves the ability to infer the underlying populations.

The rest of this paper is structured as follows: Section \ref{sec:methods} discusses the dataset which we work with and explains the standard contrast-based methods and the novel data-driven methods we use. Section \ref{sec:results} presents the prediction performance of all models, analyzes the impact of incorporating dynamical information from Rabi oscillations, and investigates how performance varies across different output states and different types of quantum circuits. For the data-driven models, we also have a look at how the prediction quality changes for different training set sizes. Finally, Section \ref{section:discussion} discusses the main findings, including the benefit of incorporating dynamical information from Rabi signals, the superior performance of data-driven regression methods over conventional and calibration-based approaches, and the observation that simple linear models already capture most of the relevant structure in the measurement signals.

\section{Methods}
\label{sec:methods}

\subsection{Experimental setup and data acquisition}
Our experiments were performed at room temperature, at different times during a two-month period. All experiments were done using a two-qubit system consisting of the electron spin and the nuclear spin of the same $N^{14}$ within the synthetic diamond. We did not perform optimal control. The used hardware setup is an XQ1 system manufactured by XeedQ GmbH.

In order to read out the population of each basis state of the two-qubit system, we exploit the presence of an external magnetic field of roughly $B=500\,\text{G}$, at which the NV center exhibits an \textit{excited-state level anti-crossing} \cite{FUCHS_2008, JACQUES_2009, HE_2024}. At this field, hyperfine interactions between the electron spin and the nuclear spin enable flip-flop processes between the $\ket{00}$ and the $\ket{11}$ states, which result in the $\ket{00}$ showing measurably higher photoluminescence than the other three states \cite{IVADY_2021, RONG_2015, HE_2024}. The populations of the four basis states are measured separately, by preparing the quantum state of interest, then applying specific pulses that map each basis state's population onto the $\ket{00}$ state's population, and finally estimating the $\ket{00}$ state's population via photoluminescence measurements. 

For each experiment, we define a quantum circuit to be applied to the two-qubit system. The circuits contain state initialization, then a set of up to 4 gates, including up to 2 conditional $X$-gates, and single-qubit $X$-rotation gates with arbitrary rotation angles.

After readout, we obtain a signal vector $\mathbf{x}$ of dimension $4(4+N_{Rabi})$, and will discuss the definition of this vector in the following. The full vector can be thought of as consisting of 4 blocks of length $4+N_{Rabi}$, where each block corresponds to measuring the population of one of the four basis states of the system, $\ket{00}$, $\ket{01}$, $\ket{10}$ and $\ket{11}$. For each of the $4+N_{Rabi}$ values of each of these readout blocks, different operations are performed on the two-qubit system and then a readout projection operation is performed, where the population of the basis state of interest is mapped onto the population of the $\ket{00}$ state, whose population is then read out.

The first 4 values of each block are obtained by initializing the system in each of the four basis states by the use of optical pumping techniques \cite{JACQUES_2009} and specific gate sequences, and then reading out the population of the corresponding state. These 4 values can be seen as columns of a confusion matrix or calibration matrix $C_{ij}=P(\text{measure }i|\text{prepare }j)$, i.e. the probability of measuring the state $i$ after one initialized the system in the state $j$. In the ideal case of perfect state preparation and readout, this matrix would be the identity. Deviations from this ideal behavior therefore directly quantify combined state preparation and measurement (SPAM) errors arising from imperfect initialization, decoherence and state-dependent readout infidelities. We will call these first 4 values in each readout block $i$ the calibration signal $\mathbf{c}^{(i)}$.

While the calibration values of each readout block are independent of the circuit, the last $N_{Rabi}$ values of each readout block encode the population of the readout state after the circuit is applied to the two-qubit system. The first of these $N_{Rabi}$ values is obtained by applying the circuit and the corresponding readout operation on the qubit, and then measuring the photon counts. The next value is obtained by doing the same, but applying a microwave pulse with the Rabi frequency of the system before readout, with a duration of $\omega_{Rabi}/N_{Rabi}$, with the Rabi period $\omega_{Rabi}$. The next value is then obtained by applying a Rabi pulse for a duration of $2\omega_{Rabi}/N_{Rabi}$ and so on. We will call these $N_{Rabi}$ values the Rabi signal $\mathbf{r}^{(i)}$ for each readout state $i$. 

Each value is obtained by repeating the same experiment for 500,000 sweeps, yielding a time-resolved photon trace. This photon-trace is integrated over a predefined time window for each, and the final value is the sum of these intensities.

\subsection{Ideal theoretical populations and problem formulation}
The main goal of this work is to develop models that can make good predictions $\hat{\mathbf{p}}$ of the ``true'' populations of the four computational basis states of our system, which are calculated from experimentally measured signal vectors $\mathbf{x}$. To compare the prediction quality of our models, and also to train them if the model architecture allows this, we use a reference for the `true'' populations that would be obtained in the absence of noise and gate imperfections. We obtain these ideal theoretical populations $\mathbf{p}^*$ by simulating the circuits that we apply to our physical two-qubit system with Qiskit \cite{qiskit2024}, and interpreting the relative measurement frequency of each state as our ideal theoretical population. The goal is then to find a model whose predictions $\hat{\mathbf{p}}$ are as similar as possible to the ideal simulated populations $\mathbf{p}^*$, where similarity has to be defined via a suitable metric, which will be discussed below.

\subsection{Reconstruction methods}

\subsubsection{Contrast method}
The standard method \cite{HE_2024, GRUBER_1997} to reconstruct the state populations from the photon count signals is what we will call the "contrast method". Here, for each readout state $i$, we define a range in which the readout intensity $I_i$ corresponding to the population of the readout state should lie in. The calculation of the intensities $I_i$ from the measurement signal vectors $\mathbf{x}$ will be discussed in Section \ref{sec:single_vs_dynamical}. Ideally, the upper limit of this range is given by $\mathbf{c}^{(i)}_i$, i.e. the intensity measured when the two-qubit state was initialized in the $i$-th state, and then the population of the same state is read out, which should result in an intensity corresponding to the system being purely in state $i$. The lower limit is approximated by the average of the other calibration values, where we initialize state $i$ and measure the populations of the states $j\neq i$. The population of the readout state is then calculated by linearly rescaling the readout intensity within the scale of the upper and lower bounds.

The contrast method calculates the population $\mathbf{y}_i$\footnote{Note that in this paper $\textbf{y}$ refers to the unnormalized outputs of our models, and $\hat{\mathbf{p}}$ refers to the populations after projection on to the probability simplex.} of the $i$-th readout state as 


\begin{equation}
\mathbf{y}_i(I_i) = \frac{I_{i}-l_i}{u_i-l_i},
\label{eq:basic-contrast}
\end{equation}

where

\[
l_i\equiv\frac{\sum_{j\neq i}\mathbf{c}_j^{(i)}}{3},
\qquad
u_i\equiv\mathbf{c}_i^{(i)}.
\]

where $I_i$ is the intensity corresponding to measuring the $i$-th state's population, $\mathbf{c}_j^{(i)}$ is the calibration value corresponding to initializing the $i$-th basis state and measuring the population of the $j$-th basis state and $u_i$ and $l_i$ are the upper and lower bounds of the linear intensity scale. If $I_i$ lies outside the bounds of $l_i$ and $u_i$, it will be clipped to 0 or 1. 

The resulting populations are generally not normalized. We obtain ``physical'' populations $\hat{\mathbf{p}}$ by projecting the population vector $\mathbf{y}$ onto the 4-dimensional probability simplex (see Section \ref{sec:simplex}). The contrast method will serve as a benchmark to which we can compare the other methods.

\subsubsection{Matrix method}
As a next iteration, we attempt to make better use of the information in the calibration signal by using matrix-inversion based state reconstruction \cite{BRAVYI_2021}. Assuming that there is a linear relation between the SPAM-error-free populations $P$ of each readout state and the measured, noisy intensities, this relation is given by some calibration matrix $C$ as $I=CP$, where $I$ is the vector of measured intensities and $P$ is the vector of mitigated populations. We assume that our calibration signals $\mathbf{c}^{(i)}$ approximate the columns of this matrix $C$, and obtain an estimate of the populations via 

\begin{equation}
\mathbf{y}=C^{-1}I.
\end{equation}

To obtain physical populations, we again project this vector $\mathbf{y}$ onto the 4D probability simplex.

\subsubsection{Ridge regression}
\label{sec:ridge}
Next, we examine if data-driven approaches can improve the prediction quality. The contrast method and the matrix method make assumptions about the form of the relationship between the signal values and the populations. In essence, the contrast method proposes independent affine mappings for each readout state, where the slope and the offset are given by the bright and dark values which depend linearly on the calibration values. Clipping the populations to $[0,1]$ further restricts the expressivity of this relation. The projection onto the probability simplex is not an affine transformation, but we will disregard this for the comparison of the models for now.

The matrix method also proposes a linear relation between the signals and the populations, whereas here the matrix-form of the equation allows the population of one readout state to depend on the readout intensity of another. Furthermore, the linear relation given by the matrix depends on the calibration signals in a non-linear way, as the calibration matrix is inverted to obtain the populations. Nonetheless, the relation between the populations and the readout intensities remains linear.

A reasonable next iteration of these methods is to use a linear regression model, which can in theory model any linear relationship between the signal values and the populations \cite{ML_2006}. In this case, the relationship is given as $\mathbf{y}=W\mathbf{x}+\beta$, where $\mathbf{x}\in \mathbb{R}^{4(4+N_{Rabi})}$ is the full signal vector containing 4 readout blocks with calibration and Rabi values, $W\in\mathbb{R}^{4\times 4(4+N_{Rabi})}$ is a learnable weight matrix and $\beta\in\mathbb{R}^{4}$ is a learnable offset vector. Using our training data $\{\left( \mathbf{p}^*_i, \mathbf{x}_i\right)\}_{i=1}^N$, there is an analytical solution that minimizes the mean-squared error between the simulated populations $\mathbf{p}^*$ and the predictions $\mathbf{y}$ \cite{HASTIE_2001}.

In contrast to the contrast method and the matrix method, with a linear regression model, the solution depends on the training data, which means that we need diverse training data and to address the problem of over-fitting if we want to get a model that generalizes well, especially to unseen data. We use $\ell_2$-regularization to prevent over-fitting \cite{HASTIE_2001}, so our final \textit{ridge regression} training objective becomes

\begin{equation}
\hat{W}, \hat{\beta} = \arg\min_{W,\beta} \sum_{i=1}^N \left\| \mathbf{p}_i^* - (W \mathbf{x}_i + \beta) \right\|_2^2 \;+\; \alpha \|W\|_F^2,
\end{equation}

where $\alpha$ is the regularization parameter, whose optimal value we will find via hyperparameter optimization. The parameters $\hat{W}$ and $\hat{\beta}$ are obtained by minimizing the regularized mean-squared error in closed form. The linear regression model also does not automatically predict physical populations. Therefore, we project the linear regression model's predictions onto the probability simplex, as before.

\subsubsection{Kernel ridge regression}
Although the linear regression model and the contrast and matrix methods use a final projection onto the probability simplex, and therefore effectively represent non-linear transformations, the non-linearity only acts on the output space. The mapping from input features to pre-activation outputs remains linear. We examine whether using a non-linear model increases the prediction quality, or if the relation between the signals and the populations is effectively captured by a linear model.

To this end, we consider kernel ridge regression, which essentially extends ridge regression by applying a non-linear transformation of the input data via a kernel function \cite{RASMUSSEN_2006}. Instead of explicitly mapping the input vectors into a higher-dimensional feature space, the model relies on pairwise similarities between data points.

Given an input vector $\mathbf{x}$, we define the kernel vector $\mathbf{k}(\mathbf{x}) \in \mathbb{R}^N$ with entries

\begin{equation}
\mathbf{k}_i(\mathbf{x}) = k(\mathbf{x}, \mathbf{x}_i),
\end{equation}

where $\{\mathbf{x}_i\}_{i=1}^N$ are the training samples and $k(\cdot,\cdot)$ is a kernel function. The prediction is then given by

\begin{equation}
\mathbf{y}(\mathbf{x}) = \mathbf{k}(\mathbf{x})^\top A + \beta,
\end{equation}

where $A \in \mathbb{R}^{N \times 4}$ is the \textit{dual coefficient matrix}, similar to the weight matrix $W$ in Section \ref{sec:ridge}, and $\beta \in \mathbb{R}^4$ is an offset vector.

We use the Radial Basis Function (RBF) kernel \cite{ML_2006}
\begin{equation}
k(\mathbf{x}, \mathbf{x}_i) = \exp\left(-\gamma \|\mathbf{x} - \mathbf{x}_i\|_2^2 \right),
\end{equation}
where $\gamma > 0$ is a hyperparameter controlling the width of the kernel.

To prevent overfitting, we again employ $\ell_2$-regularization on the coefficients of the dual coefficient matrix $A$, resulting in a kernel ridge regression objective:

\begin{equation}
\hat{A}, \hat{\beta} = \arg\min_{A,\beta} \sum_{i=1}^N \left\| \mathbf{p}_i^* - (\mathbf{k}(\mathbf{x})^\top A + \beta) \right\|_2^2 \;+\; \alpha \|A\|_F^2.
\end{equation}

Both the regularization parameter $\alpha$ and the kernel parameter $\gamma$ are selected via hyperparameter optimization. The model parameters are obtained by minimizing the corresponding kernel ridge regression objective. Again, the final prediction is projected onto the probability simplex.

\subsubsection{Dynamical intensity readout}
\label{sec:single_vs_dynamical}
We look at two different methods to calculate the intensities $I_i$. The standard approach is to apply the circuit, project the readout state of interest's population onto the $\ket{00}$ state's population, and measure the photon intensity \cite{GRUBER_1997}. This measurement corresponds to $\mathbf{r}^{(i)}_0$, i.e. the first value of the Rabi signal, where no Rabi pulse is applied before readout. We will call these the \textit{single-point} intensities.

Our second approach adds information from the dynamics of the system into the intensity measurement, by fitting a sine function to the Rabi signal \cite{JELEZKO_ELECTRON_2004}, where the $N_{Rabi}$ measured intensities correspond to different values of the control parameter $\tau$, which is the pulse length. $I_i$ then corresponds to the intercept of this fit, i.e. the value of the fit function at $\tau=0$. We will call these the \textit{dynamical} intensities.

For the contrast method and the matrix method, we compare both approaches to calculate the intensities from which the populations are then calculated. 

For the linear regression and the kernel regression methods, we do not explicitly calculate the intensities via the sine fit function. For the dynamical approach, we use the full signal $\mathbf{x}\in\mathbb{R}^d$ with $d=4(4+N_{Rabi})$ as input to the models. We compare this to using a reduced version of the signal, where we use only the first of the Rabi signals and only keep $\mathbf{r}^{(i)}_0$, resulting in $\tilde{\mathbf{x}}\in\mathbb{R}^{20}$.

\subsection{Projection onto the probability simplex}
\label{sec:simplex}
The output of all reconstruction methods is a real-valued vector $\mathbf{y} \in \mathbb{R}^4$, which does not in general satisfy the constraints of the values $\mathbf{y}_i$ adding to up 1, and being non-negative, which are necessary for populations which are physically valid. We therefore project all predictions onto the 4-dimensional probability simplex

\begin{equation}
\Delta^4 = \left\{ \mathbf{v} \in \mathbb{R}^4 \,\middle|\, \mathbf{v}_i \ge 0,\; \sum_{i=1}^4 \mathbf{v}_i = 1 \right\}.
\end{equation}

We project the prediction vector $\mathbf{y}$ onto the probability simplex by finding the vector $\hat{\mathbf{p}}\in\Delta^4$ which is closest to $\mathbf{y}$ with respect to the euclidean norm. Projecting $\mathbf{y}$ onto $\Delta^4$ is therefore equivalent to finding the minimum of the optimization problem

\begin{equation}
\hat{\mathbf{p}} = \arg\min_{\mathbf{v}\in\Delta^4}||\mathbf{v}-\mathbf{y}||^2_2.
\end{equation}

\cite{SIMPLEX_2013} presents an efficient algorithm to find the analytical minimum of this optimization problem, which we use in this work to turn unnormalized predictions $\mathbf{y}$ into physically valid population vectors $\hat{\mathbf{p}}$. 

\subsection{Training procedure and evaluation}
The matrix method and the contrast method can be applied to new data as is. However, the linear regression and kernel regression methods need to be fitted to training data. The number of learnable parameters in both the linear regression and kernel regression model is not negligible compared to the size of our training set. Therefore, we can assume that the models will predict the populations of signals they have been trained on very well \cite{HASTIE_2001}, and we will probably get an overestimation of the quality of the model's predictive capabilities if we fit and evaluate them naively. 

For this reason we will fit and evaluate the linear regression and kernel regression models by doing 10-fold cross-validation \cite{HASTIE_2001}. This means that we shuffle and then split the training set into 10 batches of roughly equal size, fit a model on 9 of these batches, and then evaluate its prediction accuracy on the remaining batch.

For evaluation of the prediction quality of our models, we will use the fidelity defined as 

\begin{equation}
F(\mathbf{p}^*, \hat{\mathbf{p}}) = \left( \sum_{i=1}^4 \sqrt{\mathbf{p}^*_i  \hat{\mathbf{p}}}_i \right)^2,
\end{equation}

which corresponds to the squared Bhattacharyya coefficient \cite{BHATTA_1946}, as our main metric. To check whether this is consistent, we will also evaluate the mean-squared error and the total variation distance between the predicted and true populations.

\section{Results}
\label{sec:results}

\subsection{Reconstruction of state populations}

\begin{table*}[ht]
\centering
\begin{tabular}{lccc}
\hline
Method & Fidelity & Total Variation & MSE \\
\hline
Contrast (single-point)     & $76,9\% \pm 20,1\%$ & $0.269 \pm 0.190$ & $0.089 \pm 0.127$ \\
Contrast (dynamical)     & $79,8\% \pm 18,7\%$ & $0.250 \pm 0.184$ & $0.079 \pm 0.121$ \\

Matrix (single-point) & $76,5\% \pm 22,1\%$ & $0.270 \pm 0.208$ & $0.096 \pm 0.154$ \\
Matrix (dynamical) & $79,5\% \pm 21,4\%$ & $0.246 \pm 0.202$ & $0.085 \pm 0.153$ \\

Ridge Regression (single-point)  & $79,0\% \pm 16,1\%$ & $0.269 \pm 0.149$ & $0.070 \pm 0.099$ \\
Ridge Regression (dynamical) & $86,0\% \pm 12,7\%$ & $0.197 \pm 0.130$ & $0.042 \pm 0.076$ \\

Kernel Ridge Regression (single-point)  & $80,7\% \pm 15,2\%$ & $0.251 \pm 0.144$ & $0.062 \pm 0.072$ \\
Kernel Ridge Regression (dynamical) & $86,6\% \pm 11,8\%$ & $0.190 \pm 0.127$ & $0.039 \pm 0.057$ \\

\hline
\end{tabular}
\caption{Performance comparison of the considered state reconstruction methods over $N=497$ samples. Values are reported as mean $\pm$ standard deviation.}
\label{tab:method_comparison}
\end{table*}

\begin{figure}[ht]
\centering
\includegraphics[width=\columnwidth]{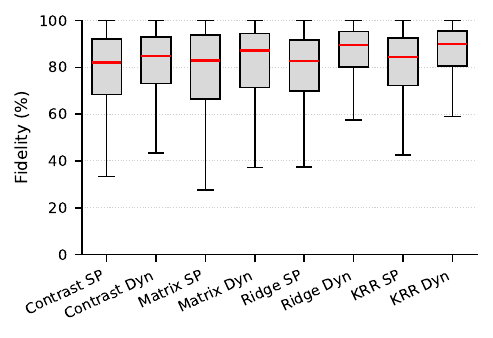}
\caption{Prediction fidelities for eight models: Contrast method, Matrix method, Ridge regression, and Kernel Ridge regression (KRR), each evaluated using single-point (SP) and dynamical (Dyn) intensities. Boxplots show the distribution of fidelities across samples, with medians indicated by the red line, interquartile ranges by the boxes, and whiskers extending to 1.5× the interquartile range.}
\label{fig:boxplot}
\end{figure}

Figure \ref{fig:boxplot} compares the distributions of the prediction fidelities for all methods, with single-point intensities and dynamical intensities. For the dynamical measurement signals, we use $N_{Rabi}=10$, and therefore $d=56$. For Ridge regression, $\alpha = 0.03$ was selected via hyperparameter optimization. Kernel Ridge regression uses $\alpha = 0.01$ and $\gamma = 1$. For both regression methods, performance was evaluated using 10-fold cross-validation.

The Ridge regression model with dynamical intensities achieves a mean fidelity of $86\%\pm13\%$ (SD, N=491), significantly outperforming both the contrast and matrix methods, particularly when using dynamical intensities. The non-linear Kernel Ridge regression has a slightly higher mean prediction fidelity than Ridge regression for single-point and dynamical intensities. Table \ref{tab:method_comparison} shows the mean fidelity, total variation distance and mean squared error for each method. The ranking of methods is roughly consistent across metrics but not completely aligned.

\subsection{Single-point intensities vs. dynamical intensities}

\begin{figure}[h]
\centering
\includegraphics[width=\columnwidth]{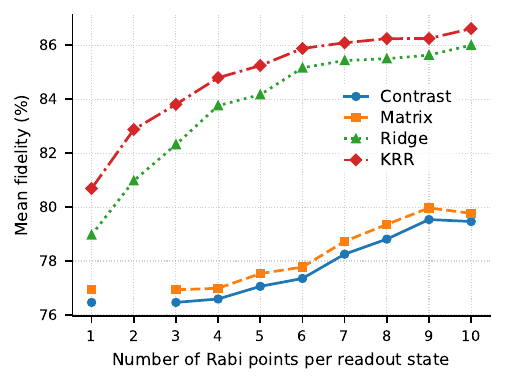}
\caption{Prediction fidelity as a function of the number of Rabi points used as input. For the contrast and matrix methods, a sine fit was performed on the first $n$ points of the Rabi signal, with the fitted intercept taken as the intensity. For the regression models, the first $n$ points were provided as input features to the model for training and inference. The fidelities were obtained through 10-fold cross-validation.}
\label{fig:rabi_points_error}
\end{figure}

Figure \ref{fig:boxplot} and table \ref{tab:method_comparison} both show that for all four methods, using the dynamical intensities improves the average fidelity, mean-squared error and the total variation distance. The improvement in prediction fidelity is relatively and absolutely larger for the regression methods than for the contrast and matrix methods.

Figure \ref{fig:rabi_points_error} compares the fidelities achieved by all models for various numbers of Rabi points used. Using one Rabi point corresponds to using the single-shot intensities. For the contrast and matrix method we did not use two Rabi points, as it is not possible to do a sine fit with only two points. We observe a nearly steady increase in the average fidelity when using more Rabi points. The average fidelity of the contrast and matrix methods actually slightly decreases when using ten instead of nine Rabi points, whereas for the regression methods it still improves. These effects are small when compared with the standard deviations though. The standard deviations obtained from cross-validation are not plotted in order to prevent a cluttered plot.

For all further analysis, we will use the dynamical intensities unless explicitly stated otherwise.

\subsection{Further analysis}




\begin{figure}[ht]
\centering
\includegraphics[width=\linewidth]{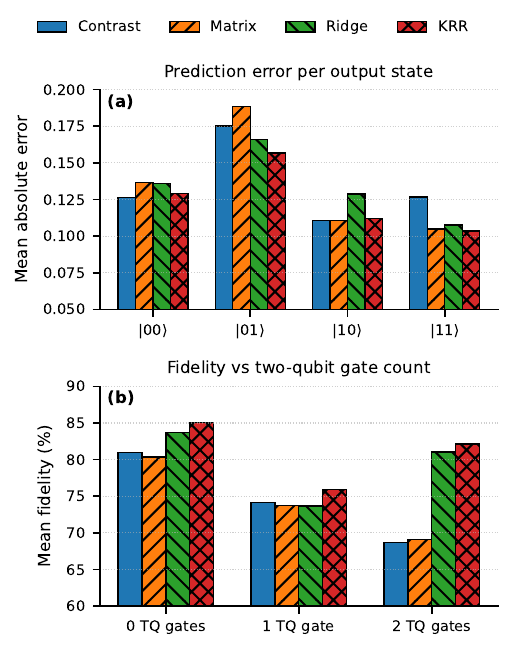}
\caption{a): comparison of mean absolute error in prediction of each states' population for each method. b): fidelity of predictions of each method considering only circuits with no two-qubit gates (left), one two-qubit gate (middle) and two-qubit gates (right). Both methods are based on the same cross-validation fidelities as before.}
\label{fig:per-state-error-two-qubit}
\end{figure}

\begin{figure}[ht]
\centering
\includegraphics[width=\linewidth]{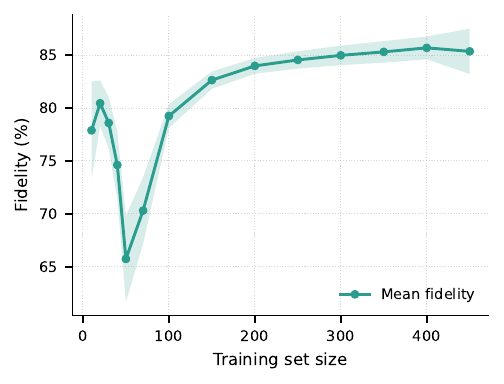}
\caption{Fidelity of linear regression predictions as a function of training set size. For each training set size, we train 10 models on randomly sampled subsets of the data and evaluate them on the remaining samples of the full dataset. The solid line shows the mean fidelity over repeated runs, while the shaded region indicates one standard deviation across repetitions.}
\label{fig:n-train-error}
\end{figure}

To better understand the behaviour of the models, we look closer at the predictions and their errors. Figure \ref{fig:per-state-error-two-qubit} a) shows the average of the absolute error for each state separately. The $\ket{01}$ state has significantly larger errors than the others across all models. The state-dependency of the errors appears consistent for all methods.

Figure \ref{fig:per-state-error-two-qubit} b) shows how the prediction error changes for data obtained with circuits including different numbers of two-qubit gates. The calibration and matrix methods show a decrease of prediction fidelity as the number of two-qubit gates increases. For the regression based models, the fidelity decreases when going from none to one two-qubit gate, but increases again when looking at circuits with two two-qubit gates gates.

To estimate if we have enough training data to train our regression models, we look at how the fidelity of the predictions changes for different training set sizes. Figure \ref{fig:n-train-error} shows the fidelity of linear regression models, which have been trained on random subsets of the entire training set for different sizes of the subsets and evaluated on the rest of the full training set. For each training set size, this procedure was repeated 10 times, and the average of these 10 fidelities is plotted. We observe a significant dip at roughly $n=50$, while $n=10, 20, 30, 40$ have higher fidelities. Interestingly, the dip occurs close to the dimensionality of the dynamical feature vector (d=56). While we do not investigate this further, it may be related to interpolation-threshold effects known from linear regression models as described in \cite{SURPRISES_2022}. For $n > 50$, the mean fidelity steadily improves. After roughly $n=300$, it begins to stagnate, with only slight increases for larger training set sizes, which are accompanied by larger standard deviations of the fidelity, as the test size becomes smaller.

\section{Discussion}
\label{section:discussion}

In this work, we compared several approaches to improve the prediction fidelity of  reading out state populations of NV center qubits at room temperature. We find that data-driven regression models significantly improve prediction fidelity over the standard contrast method and the calibration-matrix-based method. In particular, including dynamical information from Rabi oscillations consistently improves performance across all methods, indicating that time-resolved fluorescence measurements contain additional information beyond single-shot intensity estimates. We assume that this advantage can be expected for any similar state readout problem.


Among the tested models, ridge regression and kernel ridge regression achieve the highest fidelities, with only small differences between the two. This suggests that the mapping from measured readout signals to state populations is approximately linear in the chosen feature space and for our setting with shallow two qubit circuits. Future work may examine whether such non-linear effects become more important when looking at deeper circuits.

When comparing the fidelity of predictions for experiments with different numbers of two-qubit gates, the conventional methods show expected behaviour, as the fidelity decreases for higher number of two-qubit gates, which usually are a dominant source of noise. The regression-based models appear to not show this behaviour. Whether this implies robustness of the data-driven models to states with possibly higher entanglement and higher levels of noise introduced by gates is not clear from our experiments and may be further examined in future work.

Despite the improvements, several limitations remain, suggesting possible directions of future work. First, the regression models are trained using simulated noise-free populations as training targets. As real devices do not produce deterministic outcomes but instead generate inherently noisy measurement statistics, this introduces a mismatch between the training objective and the physical readout process. Therefore a natural extension of our method would be to use training targets obtained from noisy simulations, perhaps by using a noise-model which has been calibrated to resemble the real device's noise statistics.

Secondly, an important next step should be to examine how the proposed method extends to systems using more qubits and more complex circuits. An interesting question is if the mapping between the readout signals and the state populations remains approximately linear even if noise levels increase through deeper circuits. Also, it should be examined if the methods can handle the need for measuring every state separately, as the number of states increases exponentially with the number of qubits. Further work could examine the possibility of developing methods which do not need separate measurements for each basis state, to enable better scalability.

Finally, it would be interesting to compare methods on more diverse training sets. Even though our results suggest that we have an adequate amount of training data, i.e. the performance saturates and does not seem to improve much for growing training set sizes, the regression methods might benefit from being trained on more samples, especially when using more samples that come from more complex circuits.



\section*{Data Availability Statement}
The data that support the findings of this study are available from the corresponding author upon reasonable request.

\section*{Acknowledgments}
F.H. acknowledges financial support from the German National High Performance Computing (NHR) Association through NHR South-West.
We thank Priya Balasubramanian, Julian Rickert, Florian Frank, Matthias Gerster, and Gopalakrishnan Balasubramanian from XeedQ for extensive support in all hardware-related questions.

\bibliography{references}

@article{DOHERTY20131,
title = {The nitrogen-vacancy colour centre in diamond},
journal = {Physics Reports},
volume = {528},
number = {1},
pages = {1-45},
year = {2013},
note = {The nitrogen-vacancy colour centre in diamond},
issn = {0370-1573},
doi = {https://doi.org/10.1016/j.physrep.2013.02.001},
url = {https://www.sciencedirect.com/science/article/pii/S0370157313000562},
author = {Marcus W. Doherty and Neil B. Manson and Paul Delaney and Fedor Jelezko and Jörg Wrachtrup and Lloyd C.L. Hollenberg}
}

@article{CHILDRESS_REVIEW2013,
author = {Childress, Lilian and Hanson, Ronald},
year = {2013},
month = {02},
pages = {},
title = {Diamond NV centers for quantum computing and quantum networks},
volume = {38},
journal = {MRS Bulletin},
doi = {10.1557/mrs.2013.20}
}

@article{MAZE_SENSING2008,
author = {Maze, Jeronimo and Stanwix, Paul and Hodges, Jonathan and Hong, Sangjun and Taylor, J and Cappellaro, Paola and Jiang, Liang and Dutt, Gurudev and Togan, E and Zibrov, A. and Yacoby, Amir and Lukin, M},
year = {2008},
month = {11},
pages = {644-7},
title = {Nanoscale magnetic sensing with an individual electronic spin in diamond},
volume = {455},
journal = {Nature},
doi = {10.1038/nature07279}
}

@article{GAEBEL_2006,
author = {Gaebel, Torsten and Domhan, Michael and Popa, Ion and Wittmann, Christoffer and Neumann, Philipp and Jelezko, Fedor and Rabeau, J. and Stavrias, Nikolas and Greentree, Andrew and Prawer, S. and Meijer, J. and Twamley, Jason and Hemmer, Philip and Wrachtrup, Joerg},
year = {2006},
month = {05},
pages = {408-413},
title = {Room-temperature coherent coupling of single spins in diamond},
volume = {2},
journal = {Nature Physics},
doi = {10.1038/nphys318}
}

@article{JELEZKO_ELECTRON_2004,
author = {Jelezko, F. and Gaebel, T and Popa, Ion and Gruber, A},
year = {2004},
month = {03},
pages = {076401},
title = {Observation of Coherent Oscillations in a Single Electron Spin},
volume = {92},
journal = {Physical review letters},
doi = {10.1103/PhysRevLett.92.076401}
}

@article{NEUMANN_ROOMT_2008,
author = {Neumann, Philipp and Mizuochi, Norikazu and Rempp, F and Hemmer, P and Watanabe, H and Yamasaki, Satoshi and Jacques, V and Gaebel, T and Jelezko, F},
year = {2008},
month = {07},
pages = {1326-9},
title = {Multipartite Entanglement Among Single Spins in Diamond},
volume = {320},
journal = {Science (New York, N.Y.)},
doi = {10.1126/science.1157233}
}

@article{VANOORT_ODMR_2000,
author = {van Oort, Eric and Manson, Neil and Glasbeek, M},
year = {2000},
month = {11},
pages = {4385},
title = {Optically detected spin coherence of the diamond N-V centre in its triplet ground state},
volume = {21},
journal = {Journal of Physics C: Solid State Physics},
doi = {10.1088/0022-3719/21/23/020}
}

@article{PFENDER_SPECTRO_NUCLEAR_2019,
author = {Pfender, Matthias and Wang, Ping and Sumiya, Hitoshi and Onoda, Shinobu and Yang, Wen and Dasari, Durga Bhaktavatsala Rao and Neumann, Philipp and Pan, Xinyu and Isoya, Junichi and Liu, Ren-Bao},
year = {2019},
month = {02},
pages = {594},
title = {High-resolution spectroscopy of single nuclear spins via sequential weak measurements},
volume = {10},
journal = {Nature Communications},
doi = {10.1038/s41467-019-08544-z}
}

@article{GULKA_2021,
author = {Gulka, Michal and Wirtitsch, Daniel and Ivady, Viktor and Vodnik, Jelle and Hruby, Jaroslav and Magchiels, Goele and Bourgeois, Emilie and Gali, Adam and Trupke, Michael and Nesladek, Milos},
year = {2021},
month = {07},
pages = {4421},
title = {Room-temperature control and electrical readout of individual nitrogen-vacancy nuclear spins},
volume = {12},
journal = {Nature Communications},
doi = {10.1038/s41467-021-24494-x}
}

@article{STEINER_2010,
author = {Steiner, Matthias and Neumann, Philipp and Beck, Johannes and Jelezko, F.},
year = {2010},
month = {01},
pages = {35205-},
title = {Universal enhancement of the optical readout fidelity of single electron spins at nitrogen-vacancy centers in diamond},
volume = {81},
journal = {Physical Review B},
doi = {10.1103/PhysRevB.81.035205}
}

@article{JIANG_2009,
author = {Jiang, Liang and Hodges, Jonathan and Maze, Jeronimo and Maurer, P and Taylor, J and Cory, D and Hemmer, P and Yacoby, Amir and Zibrov, A. and Lukin, M},
year = {2009},
month = {10},
pages = {267-72},
title = {Repetitive Readout of a Single Electronic Spin via Quantum Logic with Nuclear Spin Ancillae},
volume = {326},
journal = {Science (New York, N.Y.)},
doi = {10.1126/science.1176496}
}

@article{
CHILDRESS_2006,
author = {L. Childress  and M. V. Gurudev Dutt  and J. M. Taylor  and A. S. Zibrov  and F. Jelezko  and J. Wrachtrup  and P. R. Hemmer  and M. D. Lukin },
title = {Coherent Dynamics of Coupled Electron and Nuclear Spin Qubits in Diamond},
journal = {Science},
volume = {314},
number = {5797},
pages = {281-285},
year = {2006},
doi = {10.1126/science.1131871},
URL = {https://www.science.org/doi/abs/10.1126/science.1131871},
eprint = {https://www.science.org/doi/pdf/10.1126/science.1131871}}

@article{ROBLEDO_2011,
author = {Robledo, Lucio and Childress, Lilian and Bernien, Hannes and Hensen, Bas and Alkemade, Paul and Hanson, Ronald},
year = {2011},
month = {09},
pages = {574-8},
title = {High-fidelity projective read-out of a solid-state spin quantum register},
volume = {477},
journal = {Nature},
doi = {10.1038/nature10401}
}

@article{SCHMUNK_2012,
author = {Schmunk, W and Gramegna, Marco and Brida, Giorgio and Degiovanni, I. and Hofer, H and Kück, Stefan and Lolli, Lapo and Paris, M and Peters, Silke and Rajteri, Mauro and Racu, Ana},
year = {2012},
month = {04},
pages = {},
title = {Photon number statistics of NV centre emission},
volume = {49},
journal = {Metrologia},
doi = {10.1088/0026-1394/49/2/S156}
}

@article{PANADERO_2024,
author = {Panadero, I. and Espinós, H. and Tsunaki, L. and Volkova, K. and Tobalina, Ander and Casanova, J. and Acedo, P. and Naydenov, B. and Puebla, R. and Torrontegui, E.},
year = {2024},
month = {07},
pages = {},
title = {Photon-emission statistics for single nitrogen-vacancy centers},
volume = {22},
journal = {Physical Review Applied},
doi = {10.1103/PhysRevApplied.22.014035}
}

@article{HE_2024,
author = {He, Jingyan and Tian, Yu and Hu, Zachary and Ye, Runchuan and Wang, Xiangyu and lu, Dawei and Xu, Nanyang},
year = {2024},
month = {05},
pages = {},
title = {Direct readout of a nitrogen-vacancy hybrid-spin quantum register in diamond by analysis of photon arrival time},
volume = {21},
journal = {Physical Review Applied},
doi = {10.1103/PhysRevApplied.21.054041}
}

@incollection{DAVIES_1994,
  author    = {Gordon Davies},
  title     = {Properties and growth of diamond},
  booktitle = {Properties and Growth of Diamond},
  year      = {1994},
  publisher = {INSPEC, the Institution of Electrical Engineers},
  address   = {London}
}

@article{MANSON_2006,
  title = {Nitrogen-vacancy center in diamond: Model of the electronic structure and associated dynamics},
  author = {Manson, N. B. and Harrison, J. P. and Sellars, M. J.},
  journal = {Phys. Rev. B},
  volume = {74},
  issue = {10},
  pages = {104303},
  numpages = {11},
  year = {2006},
  month = {Sep},
  publisher = {American Physical Society},
  doi = {10.1103/PhysRevB.74.104303},
  url = {https://link.aps.org/doi/10.1103/PhysRevB.74.104303}
}

@article{FUCHS_2008,
  title = {Excited-State Spectroscopy Using Single Spin Manipulation in Diamond},
  author = {Fuchs, G. D. and Dobrovitski, V. V. and Hanson, R. and Batra, A. and Weis, C. D. and Schenkel, T. and Awschalom, D. D.},
  journal = {Phys. Rev. Lett.},
  volume = {101},
  issue = {11},
  pages = {117601},
  numpages = {4},
  year = {2008},
  month = {Sep},
  publisher = {American Physical Society},
  doi = {10.1103/PhysRevLett.101.117601},
  url = {https://link.aps.org/doi/10.1103/PhysRevLett.101.117601}
}

@article{JACQUES_2009,
  title = {Dynamic Polarization of Single Nuclear Spins by Optical Pumping of Nitrogen-Vacancy Color Centers in Diamond at Room Temperature},
  author = {Jacques, V. and Neumann, P. and Beck, J. and Markham, M. and Twitchen, D. and Meijer, J. and Kaiser, F. and Balasubramanian, G. and Jelezko, F. and Wrachtrup, J.},
  journal = {Phys. Rev. Lett.},
  volume = {102},
  issue = {5},
  pages = {057403},
  numpages = {4},
  year = {2009},
  month = {Feb},
  publisher = {American Physical Society},
  doi = {10.1103/PhysRevLett.102.057403},
  url = {https://link.aps.org/doi/10.1103/PhysRevLett.102.057403}
}

@article{IVADY_2021,
  title = {Photoluminescence at the ground-state level anticrossing of the nitrogen-vacancy center in diamond: A comprehensive study},
  author = {Iv\'ady, Viktor and Zheng, Huijie and Wickenbrock, Arne and Bougas, Lykourgos and Chatzidrosos, Georgios and Nakamura, Kazuo and Sumiya, Hitoshi and Ohshima, Takeshi and Isoya, Junichi and Budker, Dmitry and Abrikosov, Igor A. and Gali, Adam},
  journal = {Phys. Rev. B},
  volume = {103},
  issue = {3},
  pages = {035307},
  numpages = {13},
  year = {2021},
  month = {Jan},
  publisher = {American Physical Society},
  doi = {10.1103/PhysRevB.103.035307},
  url = {https://link.aps.org/doi/10.1103/PhysRevB.103.035307}
}

@misc{qiskit2024,
      title={Quantum computing with {Q}iskit},
      author={Javadi-Abhari, Ali and Treinish, Matthew and Krsulich, Kevin and Wood, Christopher J. and Lishman, Jake and Gacon, Julien and Martiel, Simon and Nation, Paul D. and Bishop, Lev S. and Cross, Andrew W. and Johnson, Blake R. and Gambetta, Jay M.},
      year={2024},
      doi={10.48550/arXiv.2405.08810},
      eprint={2405.08810},
      archivePrefix={arXiv},
      primaryClass={quant-ph}
}

@article{GRUBER_1997,
author = {Gruber, A. and Dräbenstedt, Alexander and Tietz, C. and Ludovic, Fleury and Wrachtrup, Joerg and Borczyskowski, C.},
year = {1997},
month = {06},
pages = {2012-2014},
title = {Scanning Confocal Optical Microscopy and Magnetic Resonance on Single Defect Centers},
volume = {276},
journal = {Science},
doi = {10.1126/science.276.5321.2012}
}

@book{ML_2006,
author = {Bishop, Christopher M.},
title = {Pattern Recognition and Machine Learning (Information Science and Statistics)},
year = {2006},
isbn = {0387310738},
publisher = {Springer-Verlag},
address = {Berlin, Heidelberg}
}

@book{HASTIE_2001,
  address = {New York, NY, USA},
  author = {Hastie, Trevor and Tibshirani, Robert and Friedman, Jerome},
  publisher = {Springer New York Inc.},
  series = {Springer Series in Statistics},
  title = {The Elements of Statistical Learning},
  year = 2001
}

@book{RASMUSSEN_2006,
  author = {Rasmussen, Carl Edward and Williams, Christopher K. I.},
  publisher = {The MIT Press},
  title = {Gaussian Processes for Machine Learning},
  year = 2006
}

@article{BHATTA_1946,
 ISSN = {00364452},
 URL = {http://www.jstor.org/stable/25047882},
 author = {A. Bhattacharyya},
 journal = {Sankhyā: The Indian Journal of Statistics (1933-1960)},
 number = {4},
 pages = {401--406},
 publisher = {Springer},
 title = {On a Measure of Divergence between Two Multinomial Populations},
 urldate = {2026-04-24},
 volume = {7},
 year = {1946}
}

@article{JATTANA_2025,
author = {Jattana, Manpreet and Lippert, Thomas},
year = {2025},
month = {12},
pages = {},
title = {Predictive tracking of the NV center based on external temperature sensors},
volume = {127},
journal = {Applied Physics Letters},
doi = {10.1063/5.0292034}
}

@article{RONG_2015,
author = {Rong, Xing and Geng, Jianpei and Shi, Fazhan and Liu, Ying and Xu, Kebiao and Ma, Wenchao and Kong, Fei and Jiang, Zhen and Wu, Yang and Du, Jiangfeng},
year = {2015},
month = {06},
pages = {},
title = {Experimental fault-tolerant universal quantum gates with solid-state spins under ambient conditions},
volume = {6},
journal = {Nature Communications},
doi = {10.1038/ncomms9748}
}

@article{BRAVYI_2021,
  title = {Mitigating measurement errors in multiqubit experiments},
  author = {Bravyi, Sergey and Sheldon, Sarah and Kandala, Abhinav and Mckay, David C. and Gambetta, Jay M.},
  journal = {Phys. Rev. A},
  volume = {103},
  issue = {4},
  pages = {042605},
  numpages = {12},
  year = {2021},
  month = {Apr},
  publisher = {American Physical Society},
  doi = {10.1103/PhysRevA.103.042605},
  url = {https://link.aps.org/doi/10.1103/PhysRevA.103.042605}
}

@article{SIMPLEX_2013,
author = {Wang, Weiran and Carreira-Perpiñán, Miguel},
year = {2013},
month = {09},
pages = {},
journal = {},
title = {Projection onto the probability simplex: An efficient algorithm with a simple proof, and an application}
}

@article{SURPRISES_2022,
author = {Hastie, Trevor and Montanari, Andrea and Rosset, Saharon and Tibshirani, Ryan},
year = {2022},
month = {04},
pages = {},
title = {Surprises in high-dimensional ridgeless least squares interpolation},
volume = {50},
journal = {The Annals of Statistics},
doi = {10.1214/21-AOS2133}
}

\end{document}